\documentclass[conference,10pt,a4paper]{IEEEtran}
\usepackage{verbatim} 
\usepackage{graphicx}
\usepackage{epstopdf}
\usepackage{multirow}
\usepackage{multicol}
\usepackage{amsmath}
\usepackage{latexsym}
\usepackage{amssymb}
\usepackage{tcolorbox}
\usepackage{enumerate}
\usepackage{subcaption} % for subfigures

\newcommand{\bd}{\mathbf}
\newcommand{\ua}{^{\mathrm{a}}}
\newcommand{\ub}{^{\mathrm{b}}}
\newcommand{\uT}{^{\mathrm{T}}}
\newcommand{\uR}{^{\mathrm{R}}}
\newcommand{\dg}{^{\mathrm{o}}}

\IEEEoverridecommandlockouts
\begin{document}

\title{Observability Analysis for Fusion of Doppler Measurements in Multistatic Radar Near the Tx-Rx Baseline}
\author{\IEEEauthorblockN{Rong Yang, Leung Hong Huat Michael}\\
\IEEEauthorblockA{DSO National Laboratories\\
	12 Science Park Drive\\
	Singapore 118225\\
	Email: yrong@dso.org.sg, lhonghua@dso.org.sg}\\
\and
\IEEEauthorblockN{Yaakov Bar-Shalom}\\
\IEEEauthorblockA{Department of ECE\\
	University of Connecticut\\
	Storrs, CT 06269, USA\\
	Email: yaakov.bar-shalom@uconn.edu}
\thanks{Proc. 29th International Conference on Information Fusion, Trondheim, Norway, June 2026.} 
}
\maketitle

\begin{abstract}
This paper studies multistatic measurement fusion when a target lies within the Tx--Rx (Transmitter-Receiver) baseline ambiguity zone, with particular emphasis on configurations involving two closely spaced stationary Tx--Rx pairs. Such configurations provide overlapping detectable regions and extend the effective detection range compared with sparsely spaced multistatic systems. However, in this region, the accuracy of range and bearing measurements degrades rapidly, and Doppler measurements often remain the only reliable information source. As a result, target trajectory estimation becomes highly challenging, with observability being marginal or even completely lost.

To address this problem, the observability of target trajectories is analyzed under various conditions, enabling system designers to assess system performance in advance. A Doppler-only measurement fusion approach is then developed, employing a multiple-initial-point Maximum Likelihood (ML) nonlinear estimator for initial state estimation, followed by dynamic state updates using an Extended Kalman Filter (EKF). Simulation results are presented and shown to be consistent with the observability analysis.
\end{abstract}

%\begin{keywords}
%Multistatic system, target tracking, unknown transmitter position, UKF.
%\end{keywords}

\section{Introduction}\label{s1}
\begin{figure}[!ht]
	\begin{center}\leavevmode
		\includegraphics[width=3.4in]{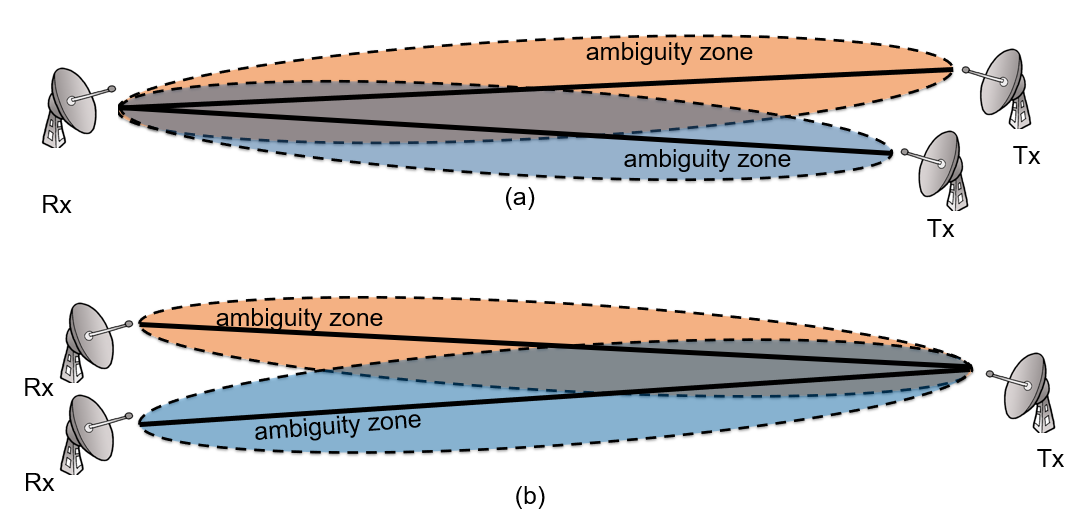}
	\end{center}
	\caption{Tx and Rx setups with overlapping ambiguity zones. (a) 2-Tx and 1-Rx. (b) 1-Tx and 2-Rx}
	\label{Fig1}
\end{figure}

Multistatic sensing systems have been widely studied due to their efficient utilization of existing transmitters. In these systems, multiple spatially separated transmitters and receivers jointly observe a moving target, and the resulting measurements are fused to estimate the target motion state. For multistatic systems, target trajectory estimation is generally considered straightforward when the target is sufficiently far from the Tx–Rx baseline, where the geometry—together with measurements of bistatic range, angle of arrival, and Doppler—is well conditioned for accurate estimation. However, when the target lies within or near the Tx–Rx baseline ambiguity zone, tracking becomes significantly more challenging. In this region, trajectory observability is marginal or may even be lost entirely, and the accuracy of range and bearing measurements deteriorates sharply. Consequently, Doppler often remains the only reliable measurement. This paper develops a Doppler-only measurement fusion approach using at least two closely spaced Tx–Rx pairs with overlapping ambiguity zones, illustrated in Fig.~\ref{Fig1}, and performs the observability analysis.

Several studies have investigated Doppler-only target tracking in multistatic configurations \cite{Maasdorp15,Battistelli13,Krysik14,Guldogan14}. In these works, the transmitters and receivers are sparsely and widely separated, resulting in sensing geometries where the bistatic baseline ambiguity regions do not overlap. Under such configurations, target observability is inherently favorable, since the target remains outside the critical near-baseline zones for all bistatic pairs. Although these studies formulate the tracking problem using Doppler-only measurements, this does not imply that Doppler is the only physically available measurement. In practice, additional information such as the bistatic range or bearing may be available in these geometries. Consequently, the tracking problem considered in these works is significantly better conditioned than the scenario addressed in this paper, where multiple Tx–Rx pairs share overlapping near-baseline ambiguity regions, leading to severely degraded observability and making Doppler-only tracking substantially more challenging.

Doppler-only tracking and localization from a single sensor have also been studied under aided conditions. A stationary target can be localized by a moving monostatic sensor \cite{Peled21}. As the sensor moves, it provides multiple, time-separated perspectives of the target, producing a multi-sensor effect that makes the problem observable without range or bearing information. Another approach uses road constraints \cite{Kim22}. When a target moves along a road, a nearby stationary sensor can track it using Doppler-only measurements, with observability ensured by the road constraint.

Our work focuses on the fusion of Doppler-only measurements from at least two closely spaced stationary Tx–Rx pairs. This configuration ensures overlapping detectable regions and provides a longer effective detection range compared with a sparsely spaced multistatic system, thanks to geometry that enhances the received signal strength. Doppler remains the only reliable measurement in this setup. This problem has attracted considerable attention in the signal processing community. In a dual-baseline bistatic radar setup, one approach estimates target velocity from two Tx–Rx pairs, assuming the target’s crossing point on the baseline is known \cite{Ustalli20}. Another method exploits the instances when a moving target crosses multiple baselines to estimate trajectory parameters, assuming these crossing times can be measured accurately \cite{Zheng22}. Both approaches rely on special assumptions and idealized conditions, and their robustness under real-world scenarios remains to be validated.

In this paper, we develop a robust dynamic estimation approach for the problem. Doppler-only tracking in multistatic radar will be applied with a Maximum Likelihood (ML) track initiation method. As this highly nonlinear problem is sensitive to initial conditions, multiple initial points are investigated for ML estimator. The paper also focuses on observability analysis under different conditions, allowing system designers to understand performance beforehand. The rest of the paper is organized as follows. Section~2 formulates the problem as a Doppler-only dynamic estimation problem. Section~3 develops the track initiation method. Section~4 analyzes the observability. Section~5 presents simulation results, and Section~6 concludes the paper.

\section{Doppler-Only Measurement Fusion}\label{s2}

To obtain a target trajectory from fusion of measurements, we formulate the problem as a dynamic estimation problem with state vector $\bd x$ 
\begin{equation}
    \bd x(k)=[x(k)\ \ y(k)\ \ \dot{x}(k)\ \ \dot{y}(k)]'
\end{equation}
where $k$ is the time index, $x(k)$, $y(k)$, $\dot{x}(k)$ and $\dot{y}(k)$ are the target position and velocity components along the $x$- and $y$-coordinates, respectively. We assume the target is moving with a nearly constant velocity (NCV) motion. The state transition model~\cite{BarShalom11} is
\begin{equation}\label{Eq_cv}
\bd x(k)=\bd F\bd x(k-1)+\bd{\Gamma}\bd v(k-1)
\end{equation}
where
\begin{eqnarray}
\bd F&=&\left[\begin{array}{cccc}
1 & 0 & T & 0\\
0 & 1 & 0 & T\\
0 & 0 & 1 & 0\\
0 & 0 & 0 & 1
\end{array}
\right]
\label{eqF}\\
\bd{\Gamma}&=&\left[\begin{array}{ccc}
0.5T^2 & 0\\
0 & 0.5T^2\\
T & 0\\
0 & T
\end{array}\right]
\label{eqG}
\end{eqnarray}
are the state transition matrix and process noise gain matrix with time interval $T$, respectively, ${\bd v}$ is zero-mean white Gaussian process noise with covariance ${\bd q}$, which is 
\begin{equation}
    \bd q(k-1) = \rm{diag}(\sigma^2_{\ddot{x}}\ \ \sigma^2_{\ddot{y}})
\end{equation}	
with $\sigma^2_{\ddot{x}}$ and $\sigma^2_{\ddot{y}}$ denoting the variances of small target acceleration errors. $\bd{\Gamma} \bd v(k-1)$ represents the state transition model error, and the model error covariance is 
\begin{equation}
\bd Q(k-1)=\bd{\Gamma} \bd q(k-1) \bd{\Gamma}'
\end{equation}

For the measurements\footnote{In the asynchronous case, $T = T(k)$ is time-varying. For simplicity, $T$ is used without explicitly showing the time argument in~(\ref{eqF})--\ref{eqG}).}, there are two cases shown in Fig.~\ref{Fig1}. In case~(a), there is a single receiver (Rx), and the measurements from the two Tx-Rx pairs are time synchronized, whereas in case~(b), the 2 Rx systems are not necessarily time synchronized. The measurement vectors are then defined as follows%, respectively.
\begin{eqnarray}
    \bd z\ua(k)&=&[\dot{r}_1(k)\ \ \dot{r}_2(k)]'\\
    z\ub_i(k)&=&\dot{r}_i(k) \hspace{10mm}i=1,2\label{eqMeas1}
\end{eqnarray}
where $\dot{r}_1(k)$ and $\dot{r}_2(k)$ are the Doppler measurements from Tx–Rx pairs 1 and 2, respectively, and $i \in \{1, 2\}$ denotes the pair index in (\ref{eqMeas1}), depending on which Tx–Rx pair provides the measurement. The measurement models can be formed using $\dot{r}_i(k)$ given below
\begin{eqnarray}
    \dot{r}_i(k) &=& h_i\big(x(k)\big) + w_i(k) \nonumber \\
    &=&\dot{r}_i\uT(k)+\dot{r}_i\uR(k) + w_i(k) \label{eqMeas2}
\end{eqnarray}
where $w_i(k)$ are  white zero-mean Gaussian measurement noises with variance $R(k)$ and
\begin{eqnarray}
    \dot{r}_i\uT(k)&=& \frac{[x(k)-x_i\uT]\dot{x}(k)+[y(k)-y_i\uT]\dot{y}(k)}{\sqrt{[x(k)-x_i\uT]^2+[y(k)-y_i\uT]^2}}\\
    \dot{r}_i\uR(k)&=& \frac{[x(k)-x_i\uR]\dot{x}(k)+[y(k)-y_i\uR]\dot{y}(k)}{\sqrt{[x(k)-x_i\uR]^2+[y(k)-y_i\uR]^2}}
\end{eqnarray}
and ($x_i\uT$,$y_i\uT$) and ($x_i\uR$,$y_i\uR$) are the Tx and Rx positions of the $i$th pair, respectively. 

Since the measurement models are nonlinear, a nonlinear state estimation approach is required. In this work, the Extended Kalman Filter (EKF) is adopted to estimate the target state.

\section{Track Initiation}\label{s3}

\begin{figure}[!ht]
	\begin{center}\leavevmode
		\includegraphics[width=2.5in]{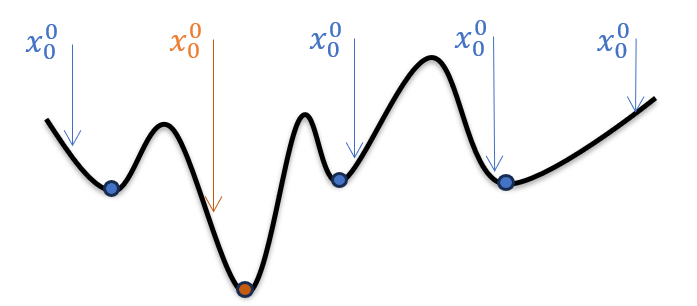}
	\end{center}
	\caption{Different starting points lead to different optima. The red dot is the global optimum and the blue dots are local optima.}
	\label{Fig2}
\end{figure}

This section addresses the track initiation problem. In general, an Iterative Least Squares (ILS) maximum likelihood (ML) estimator can be applied to an initial batch of measurements to estimate the initial state $\bd x_0$. The estimator requires an initial guess $\bd x_0^0$, from which the estimate is iteratively updated as $\bd x_0^1$, $\bd x_0^2$, and so on. However, in the considered Doppler-only measurement scenario, it is difficult to obtain a reliable initial guess $\bd x_0^0$. Fig.~\ref{Fig2} illustrates the importance of the initial guess in the track initiation process. This section first formulates the track initiation problem as an ILS estimation problem and then discusses the choice of the initial guess.

The initial state to be estimated is defined as
\begin{equation}
    \bd x_0=[x(0)\ \ y(0)\ \ \dot{x}\ \ \dot{y}]'
\end{equation}

The batch of initial measurements is given by
\begin{equation}\label{eqMeasbt}
    \bd Z=[\dot{r}_1(n_{1s})\ \ \dots\ \ \dot{r}_1(n_{1e})\ \ \dot{r}_2(n_{2s})\ \ \dots\ \ \dot{r}_2(n_{2e})]'
\end{equation}
where ($n_{1s}$, $n_{1e}$) and ($n_{2s}$, $n_{2e}$) denote the first and last time indices within the predefined initialization batch interval $[0, T_{\mathrm{init}}]$ for sensor pairs~1 and~2, respectively. Under the constant-velocity motion assumption and using the measurement model in~(\ref{eqMeas2}), the batch measurement vector (of dimension $n_{\bd Z}$) can be written as
\begin{eqnarray}
    \bd Z &=& \big[h_1\big(x(n_{1s})\big)\ \ \dots\ \ h_1\big(x(n_{1e})\big)\nonumber\\
    &&h_2\big(x(n_{2s})\big)\ \ \dots\ \ h_2\big(x(n_{2e})\big)\big]'+\bd w_{\bd Z}\nonumber\\
    &=&\bd g\big(\bd x_0\big) + \bd w_{\bd Z}
\end{eqnarray}
where $\bd w_{\bd Z}$ is the stacked measurement noise corresponding to $w$ in~(\ref{eqMeas2}), and $\bd g(\cdot)$ denotes the nonlinear mapping from the initial state $\bd x_0$ to the measurements under the constant-velocity motion model.

The initial state can be estimated by the Iterated Least Squares (ILS) estimator\footnote{We choose the ILS as the implementation of the Maximum Likelihood Estimator (MLE) \cite{B7}.}
iteratively given by
\begin{eqnarray}\label{eqILS}
\hat{\bd x}_0^{j+1}
&=&\hat{\bd x}_0^j+\left(
\bd J'(\hat{\bd x}_0^j)\bd R_{\bd Z}^{-1}\bd J(\hat{\bd x}_0^j)\right)^{-1}
\bd J'(\hat{\bd x}_0^j)\bd R_{\bd Z}^{-1}\nonumber\\
&&\times\left[\bd Z - \bd g(\hat{\bd x}_0^j)\right]
\end{eqnarray}
where $\hat{\bd x}_0^j$ denotes the estimate of the initial state at the $j$th iteration, $\bd R_{\bd Z}$ is the covariance matrix of the measurement noise $\bd w_{\bd Z}$ and 
\begin{equation}\label{eqJac}
    \bd J(\hat{\bd x}_0^j)=\left.\frac{\partial \bd g(\bd x_0)}{\partial \bd x_0}\right|_{\bd x_0=\hat{\bd x}_0^j}=\left.(\nabla_{\bd x_0}\bd g(\bd x_0)')'\right|_{\bd x_0=\hat{\bd x}_0^j}
\end{equation}
is the Jacobian matrix of $\bd g(\cdot)$ (of dimension $n_{\bd Z}\times n_{\bd x_0}$) evaluated at $\hat{\bd x}_0^j$. The details are given in Appendix~\ref{a1}.

The stopping criterion of the ILS is
\begin{equation}\label{eqStop}
\left[\bd Z - \bd g(\hat{\bd x}_0^j)\right]'
\bd R_{\bd Z}^{-1}
\left[\bd Z - \bd g(\hat{\bd x}_0^j)\right]\leq \chi^2_{n_{\bd Z}-n_{\bd x_0}}(95\%)
\end{equation}
where the threshold is the 95-th percentile of the chi-square density with $n_{\bd Z}-n_{\bd x_0}$ degrees of freedom \cite{B7}.

It is difficult to obtain a good initial guess for $\bd x_0$. The paper therefore utilizes multiple ($N_0$) initial guesses $\{\bd x_0^{0i}\}_{i=1}^{N_0}$. This is achieved by uniformly sampling positions within the detectable zone, assigning uniformly distributed omnidirectional headings, and uniformly distributed speeds in the interval $[s_{\mathrm{min}}, s_{\mathrm{max}}]$. Here, $s_{\mathrm{min}}$ is set to the maximum measured $\dot{r}_i(k)$ in $\bd Z$, since the target speed cannot be smaller than its radial component, and $s_{\mathrm{max}}$ is the predefined maximum target speed. 

For each candidate, the measurement residual is computed as
\begin{equation}
    e^i = \left[\bd Z - \bd g(\bd x_{0}^{0i})\right]'
          \left[\bd Z - \bd g(\bd x_{0}^{0i})\right]
\end{equation}
where $i$ denotes the $i$th candidate. The candidate that yields the minimum residual is then selected as the initial state estimate $\bd {\hat x}_0^0$.

\section{Observability Analysis}\label{s4}

This section presents the observability analysis of the considered problem. Observability indicates whether a target state can be uniquely determined from a finite set of measurements. This is assessed through the observation matrix, which maps the state vector to the measurements. If the mapping is a nonlinear function, the Jacobian of the function with respect to the state is used as the observation matrix. In this work, the nonlinear measurement function is defined in~(\ref{eqMeasbt}), with its Jacobian denoted by $\bd J(\bd x_0)$ in~(\ref{eqJac}) and the observability is quantified by the scalar ``observability measure" expression\footnote{The $n_\bd x \times n_\bd x$ matrix $\bd J(\bd x_0)' \,\bd J(\bd x_0)$ is, for a diagonal noise covariance matrix $\bd R$, proportional to the Fisher Information Matrix (FIM), whose invertibility guarantees the local observability of $\bd x_0$ \cite{B7}.}
\begin{equation}
    O(\bd x_0) = \log\left[\det\big(\bd J(\bd x_0)' \,\bd J(\bd x_0)\big)\right]
    \label{obs}
\end{equation}
A higher value of $O(\bd x_0)$ indicates that the measurements are more sensitive to changes in the state $\bd x_0$, making it easier to obtain an accurate state estimate. Conversely, a lower value corresponds to a marginally observable state. If $O(\mathbf{x}_0) = -\infty$, which is equivalent to $\det(C) = 0$, the state $\mathbf{x}_0$ is not completely observable, and its neighborhood is marginally observable.

\begin{figure}[!ht]
	\begin{center}\leavevmode
		\includegraphics[width=3.4in]{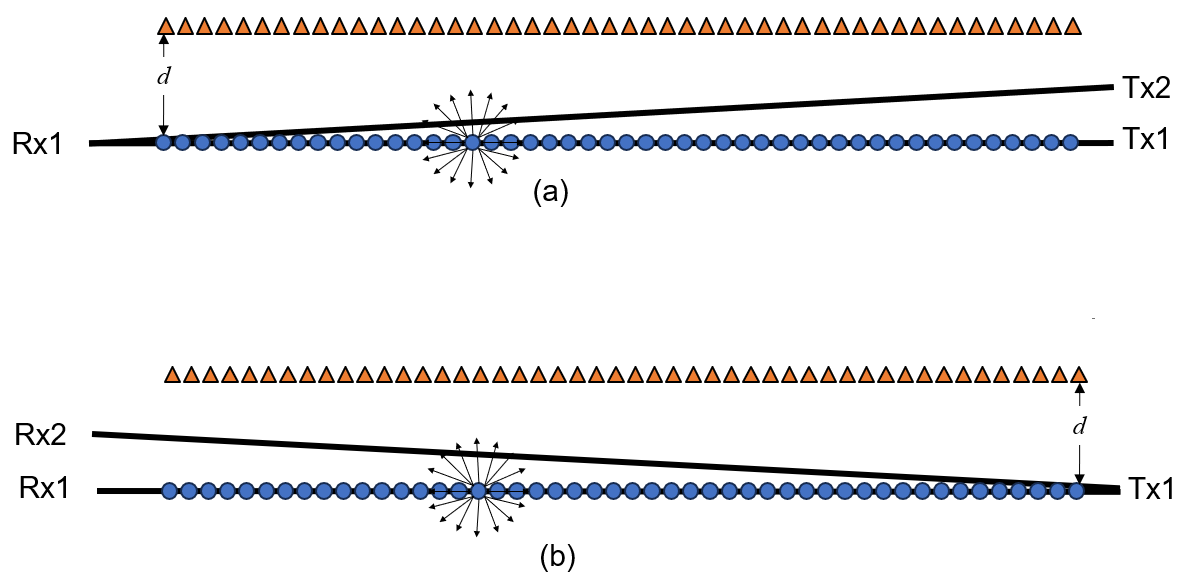}
	\end{center}
	\caption{Observability study setup. Subfigures (a) and (b) correspond to the sensor configurations in Fig.~\ref{Fig1}. The $\circ$ and $\triangle$ symbols indicate the positions of $\bd x_0$ for evaluation. Different speeds and headings are used at each point, the latter indicated in the middle $\circ$ by the ``star".}
	\label{Fig3}
\end{figure}

In view of the complexity of the FIM in the present problem, a numerical approach is adopted in this work to evaluate the observability. Fig.~\ref{Fig3} illustrates four groups of selected initial states $\bd x_0$ for the two sensor configuration cases, which correspond to those shown in Fig.~\ref{Fig1}. The detailed sensor configurations for the two cases are listed in Table~\ref{tb_1}.
\begin{table}[!ht]
    \caption{Sensor positions}
    \label{tb_1}
    \centering
    \begin{tabular}{c|cccc}
        \hline \hline
        Case & Tx1 & Tx2 & Rx1 & Rx2 \\
        \hline
        (a) & $(50\,\mathrm{km},\,0)$ & $(50\,\mathrm{km},\,1\,\mathrm{km})$ & $(0,\,0)$ & -- \\
        (b) & $(50\,\mathrm{km},\,0)$ & -- & $(0,\,0)$ & $(0,\,1\,\mathrm{km})$ \\
        \hline
    \end{tabular}
\end{table}
The four evaluation scenarios (each being a group of initial states, as illustrated in Fig.~\ref{Fig3}) are described as follows:
\begin{itemize}
\item[1.] \textbf{Blue $\circ$ in (a)}: The states are located on the Tx1--Rx1 baseline. The $x$-coordinate varies within $[1\,\mathrm{km}, 49\,\mathrm{km}]$ with a spacing of $500\,\mathrm{m}$. The heading angles at each point range from $[0^\circ, 359^\circ]$ with increments of $10^\circ$. The target speed is fixed at $100\,\mathrm{m/s}$.

\item[2.] \textbf{Red $\triangle$ in (a)}: The states are located at a distance of $2\,\mathrm{km}$ away from the Tx1--Rx1 baseline. All other settings are identical to those in scenario~1).

\item[3.] \textbf{Blue $\circ$ in (b)}: The same configuration as in scenario~1), but applied to sensor configuration~(b).

\item[4.] \textbf{Red $\triangle$ in (b)}: The same configuration as in item~2), but applied to sensor configuration~(b).
\end{itemize}

\begin{figure}[!ht]
	\begin{center}\leavevmode
		\includegraphics[width=3.4in]{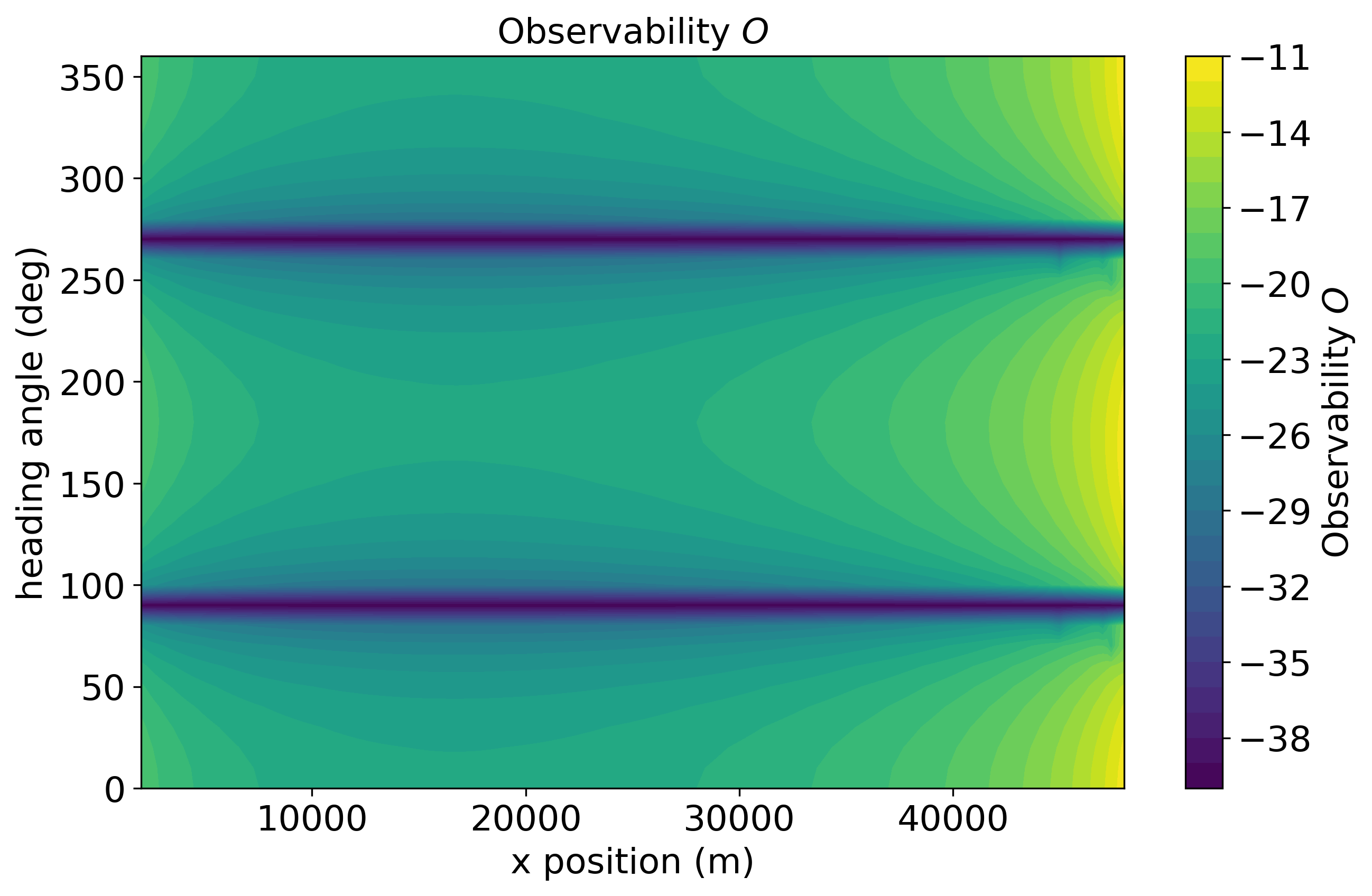}
	\end{center}
	\caption{Observability of group 1 [blue $\circ$ in Fig.~\ref{Fig3}(a)].}
	\label{Fig4}
\end{figure}
\begin{figure}[!ht]
	\begin{center}\leavevmode
		\includegraphics[width=3.4in]{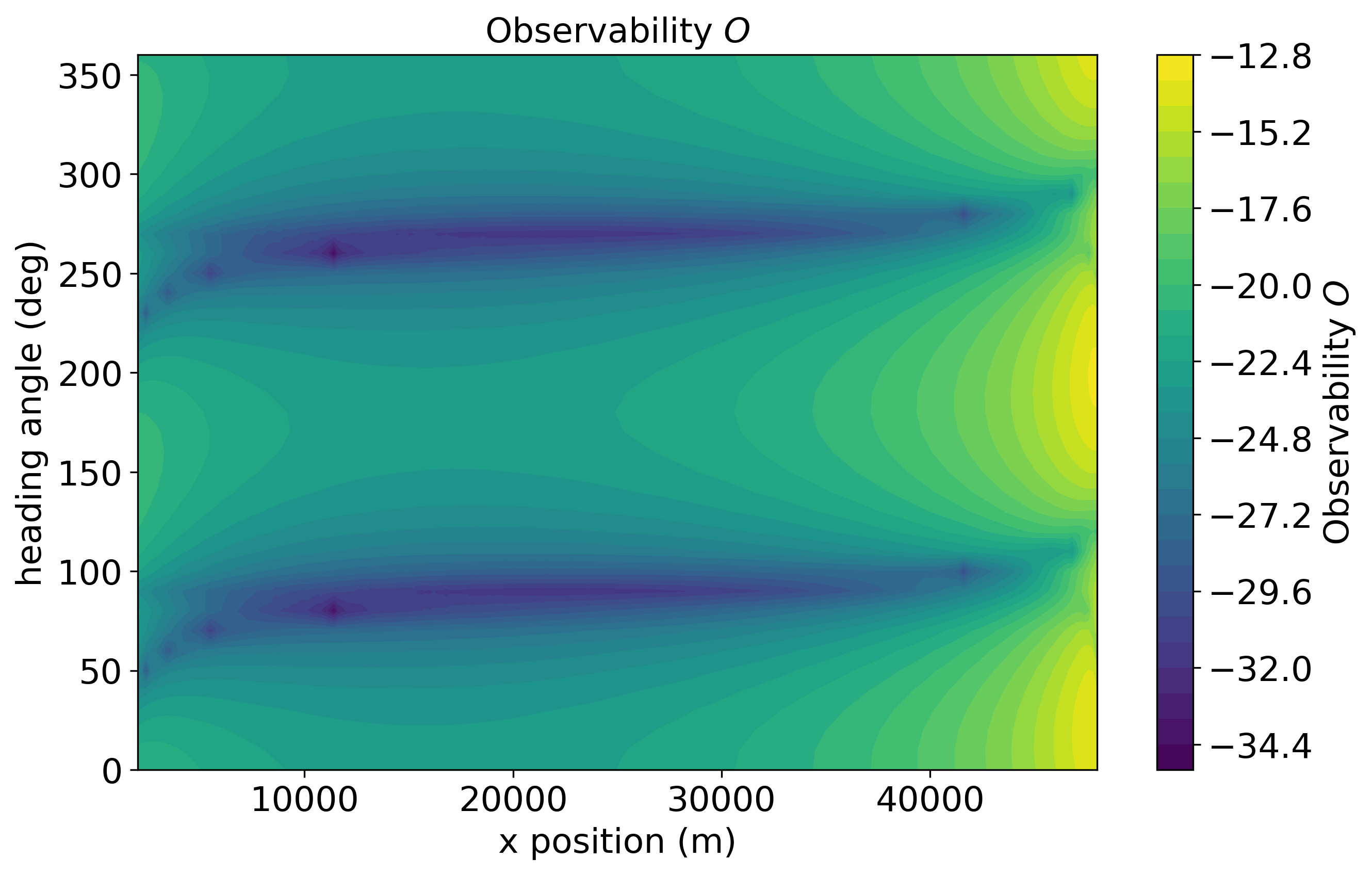}
	\end{center}
	\caption{Observability of group 2 [red $\triangle$ in Fig.~\ref{Fig3}(a)].}
	\label{Fig5}
\end{figure}
\begin{figure}[!ht]
	\begin{center}\leavevmode
		\includegraphics[width=3.4in]{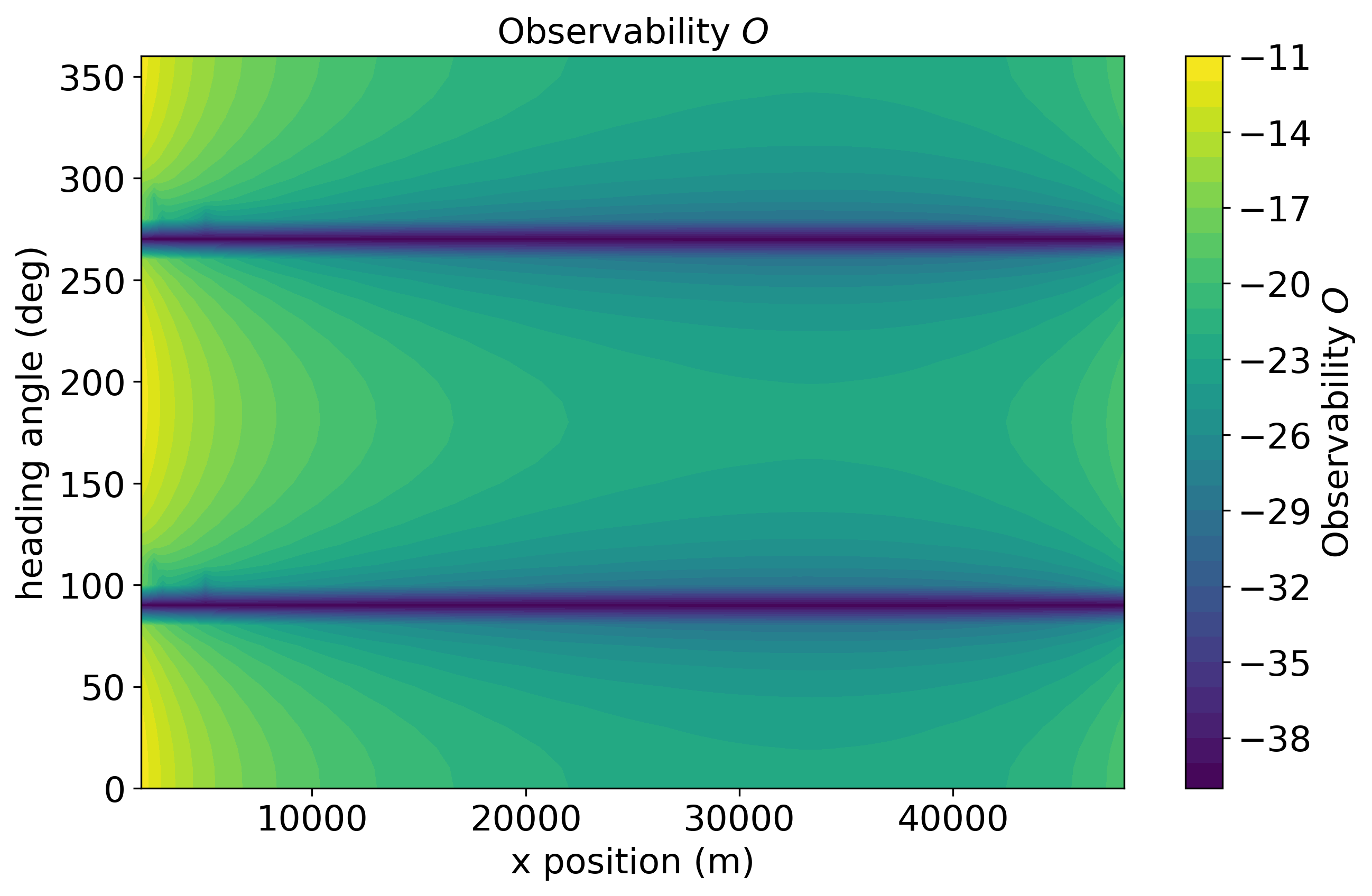}
	\end{center}
	\caption{Observability of group 3 [blue $\circ$ in Fig.~\ref{Fig3}(b)].}
	\label{Fig6}
\end{figure}
\begin{figure}[!ht]
	\begin{center}\leavevmode
		\includegraphics[width=3.4in]{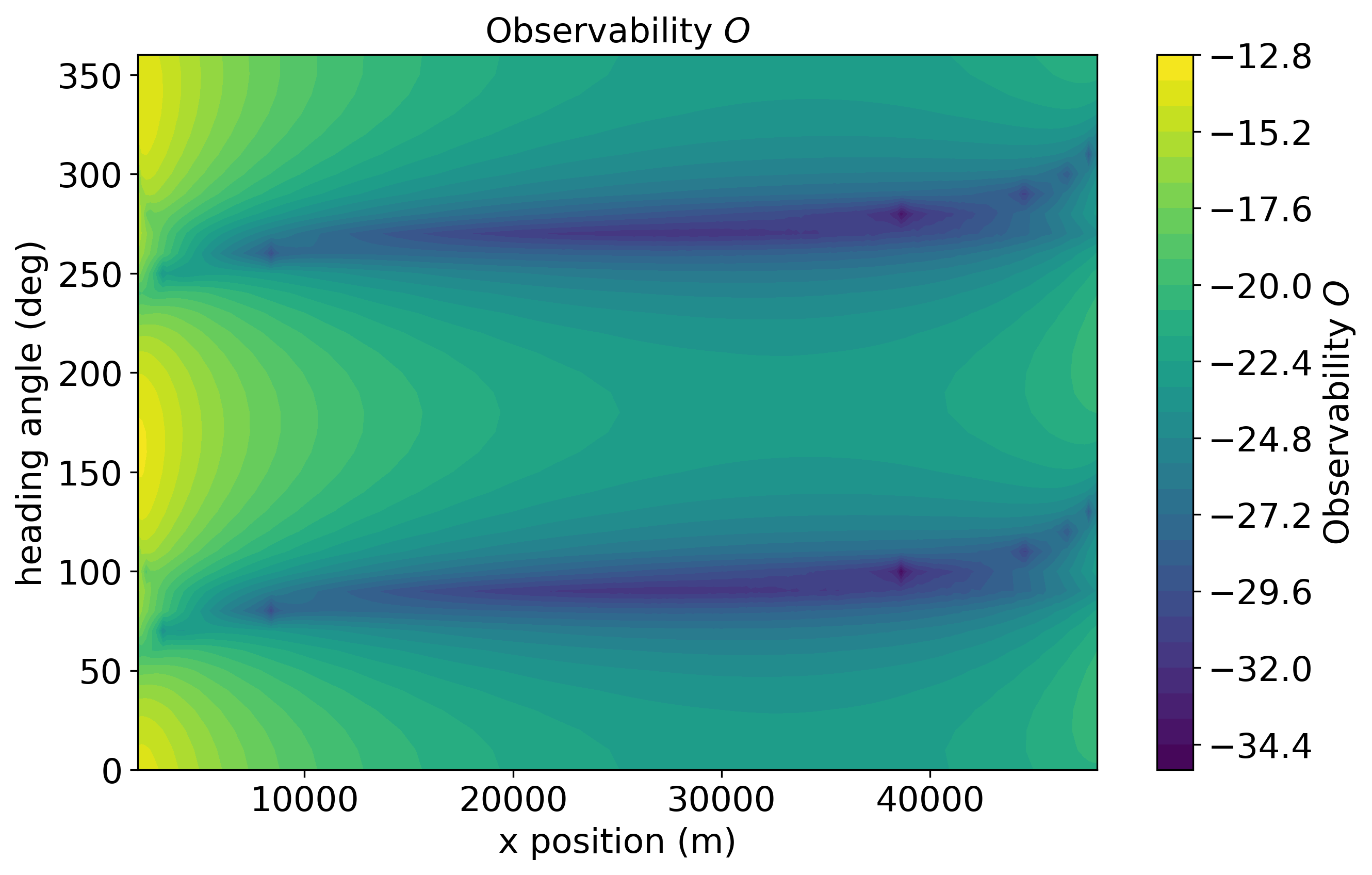}
	\end{center}
	\caption{Observability of group 4 [red $\triangle$ in Fig.~\ref{Fig3}(b)].}
	\label{Fig7}
\end{figure}

The observability values $O(\bd x_0)$ for the four evaluation groups are computed and presented in Figs.~\ref{Fig4}--\ref{Fig7}, respectively. In these figures, the horizontal and vertical axes correspond to the target initial $x$-position and heading angle, respectively. During the computation, some values of $\det(\bd J'\bd J)$ become zero due to rank deficiency of the observation matrix. To enable logarithmic visualization, these zero values are replaced by the smallest positive number representable in the computation, namely $10^{-40}$, which yields $\log(10^{-40})=-40$. Therefore, regions where $O=-40$ indicate that the corresponding state $\bd x_0$ is unobservable with respect to the $x$-position and heading angle. 

In Figs.~\ref{Fig4} and~\ref{Fig6}, groups 1 and 3 (blue $\circ$ in Fig.~\ref{Fig3}) are unobservable when the heading angles are $90^\circ$ and $270^\circ$, corresponding to the case where the target moves along the Tx1--Rx1 baseline. In this configuration, the bistatic range remains constant, and the two bistatic range rates $\dot{r}$ associated with the Tx1--Rx1 pair are zero. Since $\dot{r}$ does not change as the target moves along these directions, the observability matrix becomes singular, i.e., $\det(\bd J' \bd J)=0$.

\begin{figure}[!ht]
	\begin{center}\leavevmode
		\includegraphics[width=3.4in]{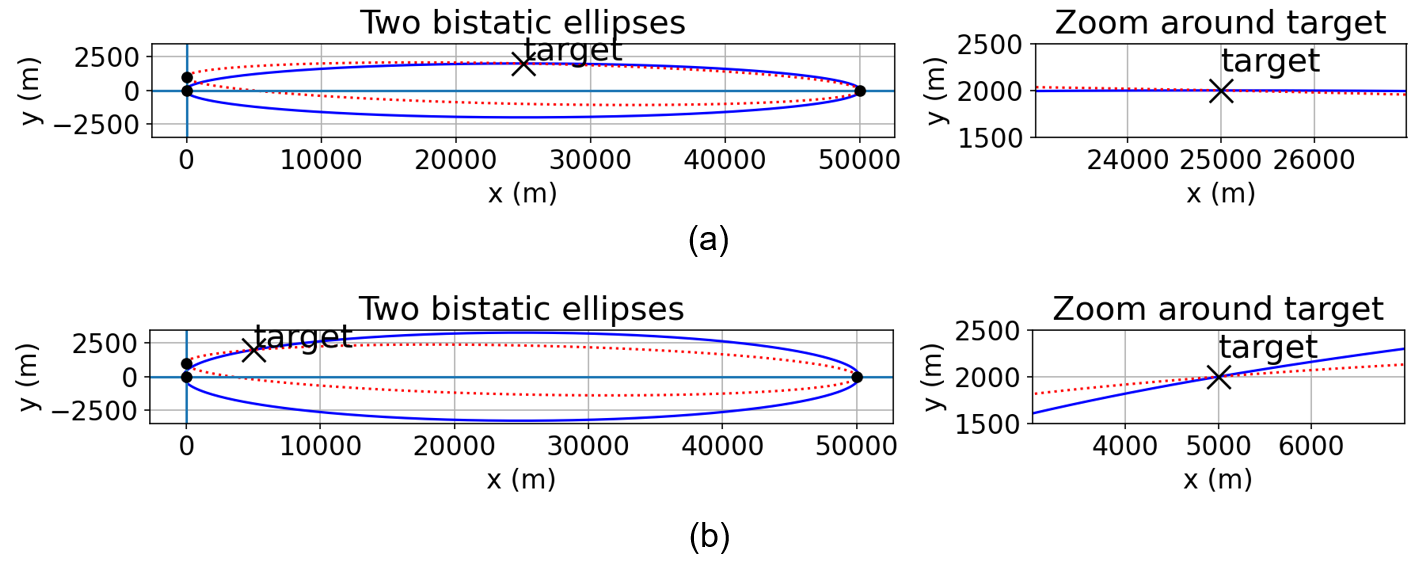}
	\end{center}
	\caption{The low observability regions. (a) at middle $x=25$km. (b) at branch $x=5$km.}
	\label{Fig7a}
\end{figure}

In Figs.~\ref{Fig5} and~\ref{Fig7} (red $\triangle$ in Fig.~\ref{Fig3}), although all observability values $O$ are greater than $-40$, two distinct regions of low observability are present. These regions are centered around heading angles of $90^\circ$ and $270^\circ$, with additional branch-like structures appearing near the beginning and end of the $x$-coordinate range. The corresponding headings align with the tangential directions of the bistatic ellipses formed by the associated Tx–Rx pairs\footnote{A bistatic ellipse is defined by a transmitter, a receiver, and a target. Any target point on the ellipse has a constant sum of distances to the transmitter and receiver. Consequently, when a target moves approximately along the ellipse, the bistatic range remains nearly constant, and the bistatic range rate $\dot{r}$ is close to zero. Since the target state $\bd x_0$ is simulated under a constant-velocity assumption, the trajectory cannot exactly follow the curvature of the ellipse; therefore, $\dot{r}$ does not reach an exact zero.}. This is illustrated in Fig~\ref{Fig7a}. At the middle of the $x$-axis, the target heading is consistent with the edges of both ellipses simultaneously, resulting in overlap in the observability map. In contrast, in the branch regions, for $x$ near 0 and 50km, only one heading can align with the edge of one ellipse at a time, while the corresponding edge of the other ellipse occurs at a different heading, producing the two separate branch structures. 

From the above analysis, the following conclusion can be drawn: a target state is unobservable if its trajectory lies along a bistatic ellipse associated with any Tx--Rx sensor pair. The Tx--Rx baseline is a special case of a bistatic ellipse with zero width. For all other target trajectories, the target state remains observable.
\begin{figure}[htbp]
    \centering
    % Subfigure (a)
    \begin{subfigure}[b]{0.4\textwidth}
        \centering
        \includegraphics[width=\textwidth]{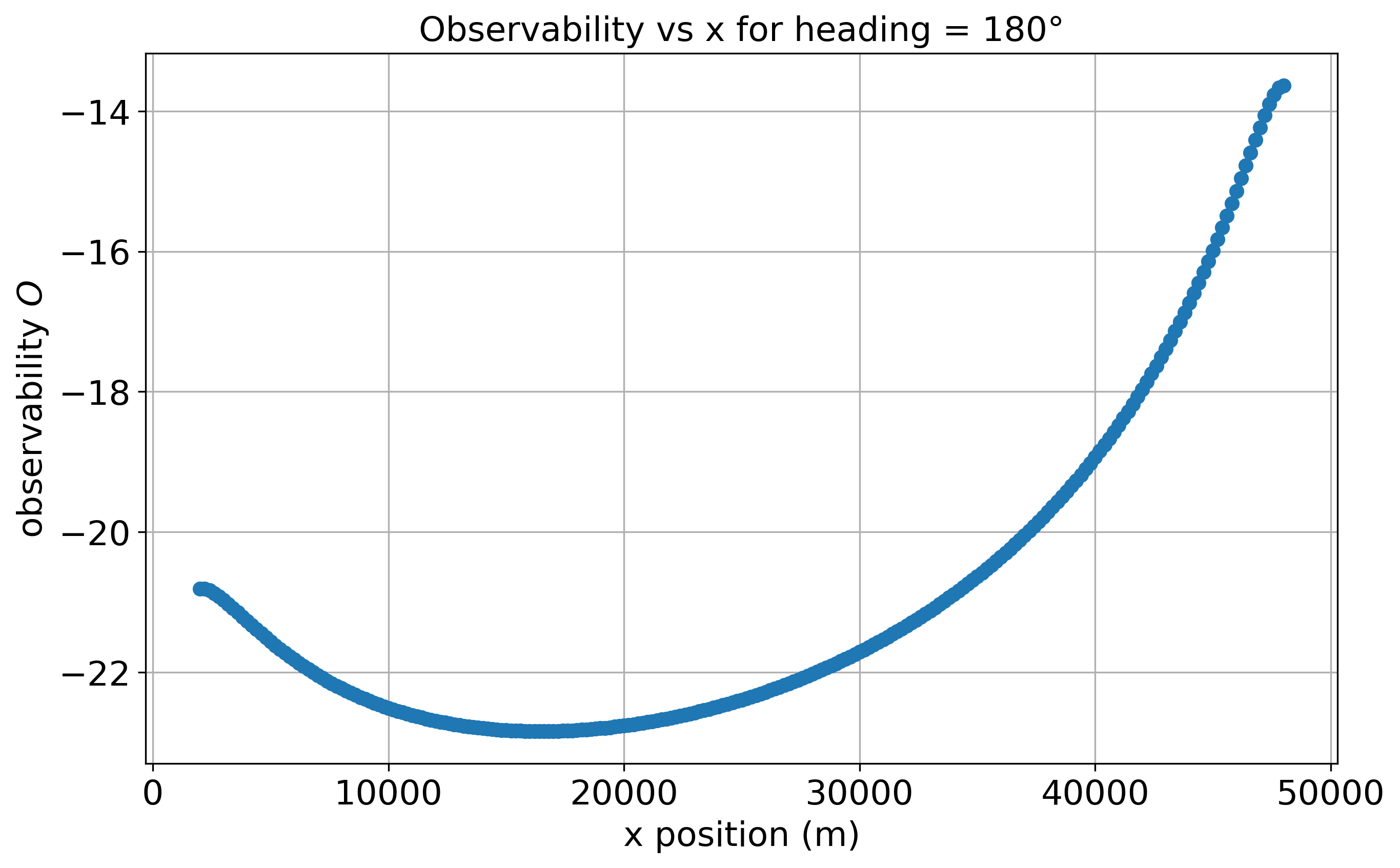}
        \caption{}
        \label{Fig8a}
    \end{subfigure}
    \hfill
    % Subfigure (b)
    \begin{subfigure}[b]{0.4\textwidth}
        \centering
        \includegraphics[width=\textwidth]{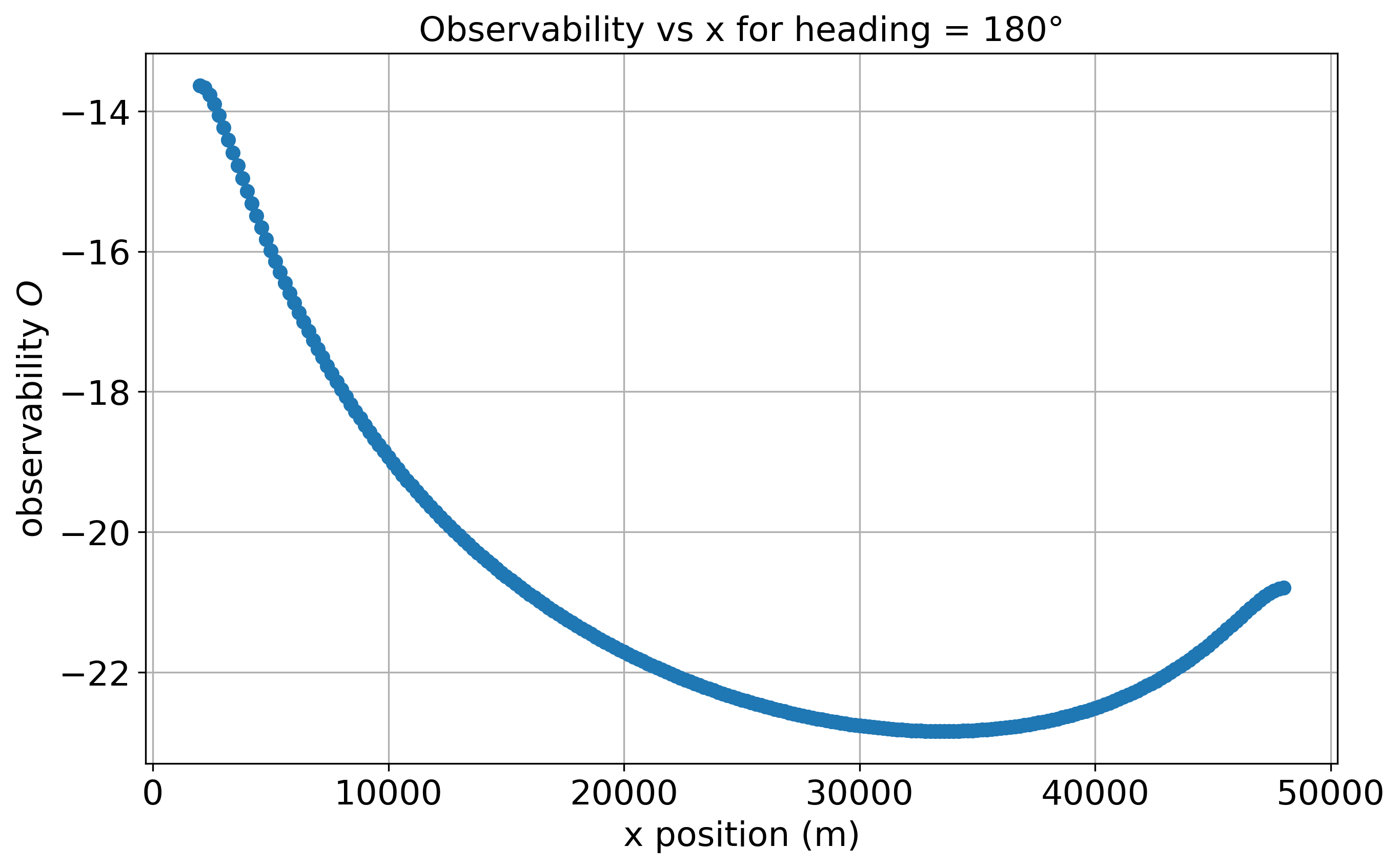}
        \caption{}
        \label{Fig8b}
    \end{subfigure}

    \caption{Observability vs.\ x for heading 180°  (a) group 2  (b) group 4.}
    \label{Fig8}
\end{figure}

To provide a detailed view of observability across different ranges, we plot the scalar observability measure (\ref{obs}) versus $x$ for a fixed heading of $180^\circ$ for groups 2 and 4 in Fig.~\ref{Fig8}. It can be seen that the target observability is lower near the middle of $x$. This occurs because the changes in $\dot{r}$ for both Tx--Rx pairs are similar in that region, whereas at the two ends of the $x$ range, the changes are more distinct. The effect is particularly pronounced when $x$ is close to 50~km in group 2 and close to 0 in group 4 in Fig.~\ref{Fig3}.

\begin{table}[!ht]
    \caption{Observability at different target speeds}
    \label{tb_2}
    \centering
    \begin{tabular}{c|cc|cc}
        \hline \hline
        Speed & \multicolumn{2}{c|}{group 2} & \multicolumn{2}{c}{group 4} \\
        & Max. & Min. & Max. & Min. \\
        \hline
        100 m/s & -10.8 & -34.9 & -10.8 & -34.9 \\
        75 m/s  & -11.9 & -34.9 & -11.9 & -34.9 \\
        50 m/s  & -13.3 & -36.3 & -13.3 & -36.5 \\
        25 m/s  & -15.7 & -38.0 & -15.7 & -38.1 \\
        \hline
    \end{tabular}
\end{table}

We also investigate the relationship between target speed and observability. Table~\ref{tb_2} lists the minimum and maximum observability values $O$ for groups~2 and~4. It is evident that the observability increases as the target speed increases.

\section{Simulation Results}\label{s5}

\begin{figure}[!ht]
	\begin{center}\leavevmode
		\includegraphics[width=3.0in]{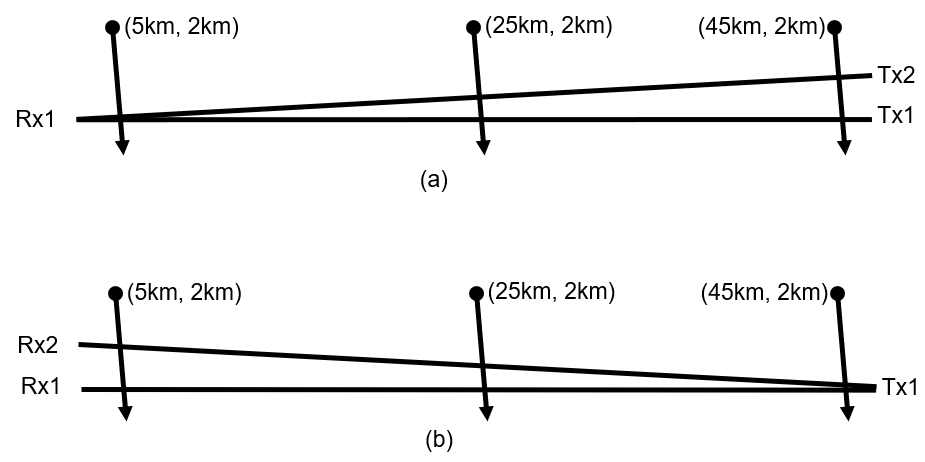}
	\end{center}
	\caption{Test scenarios.}
	\label{Fig9}
\end{figure}

Simulation tests are conducted for the scenarios shown in Fig.~\ref{Fig9}, where cases (a) and (b) correspond to the two sensor configurations given in Table~\ref{tb_1}, and three target trajectories are considered for each configuration with starting points at $(5\,\mathrm{km},\,2\,\mathrm{km})$, $(25\,\mathrm{km},\,2\,\mathrm{km})$, and $(45\,\mathrm{km},\,2\,\mathrm{km})$.
All targets share the same velocity $(10\,\mathrm{m/s},\,-100\,\mathrm{m/s})$.
The simulation duration is $30\,\mathrm{s}$ with a sampling interval of $0.5\,\mathrm{s}$.
The standard deviations of the Doppler measurement errors for the two Tx--Rx pairs are $1\,\mathrm{m/s}$.

In track initiation, the batch size is 20, with 10 measurements from each sensor pair. Multiple initial states $\bd x_0^{0,i}$, $i = 1, \ldots, N_0$, are uniformly sampled within the following ranges:
\begin{eqnarray}
    x_0 &\in& \{2000\ \ \ldots\ \ 48000\},\ \ \ \textrm{interval 1000\,m}\\
    y_0 &\in& \{-5000\ \ \ldots\ \ 5000\},\ \ \textrm{interval 1000\,m}\\
    \dot x &=& v\sin(h)\\
    \dot y &=& v\sin(h)\\
    v &\in& \{\dot r^{max}\ \ \ldots\ \ 200m/s\},\ \ \ \textrm{interval 10\,m/s}\\
    h &\in& \{0\dg\ \ \ldots\ \ 259\dg\},\ \ \ \textrm{interval 20}\dg
\end{eqnarray}
The candidate state with the minimum measurement error is selected as $\hat{\bd x}_0^0$.

\begin{figure}[!ht]
	\begin{center}\leavevmode
		\includegraphics[width=3.0in]{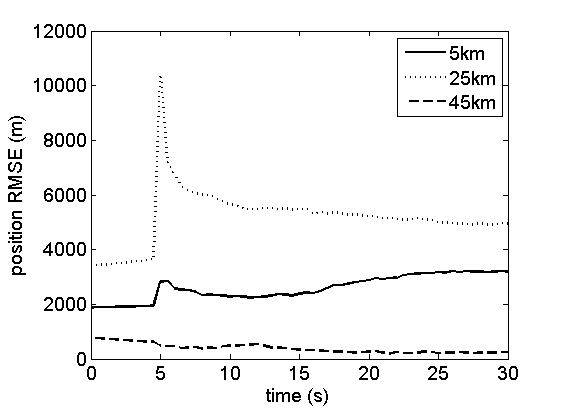}
	\end{center}
	\caption{Position RMSE for the case (a).}
	\label{Fig10}
\end{figure}
\begin{figure}[!ht]
	\begin{center}\leavevmode
		\includegraphics[width=3.0in]{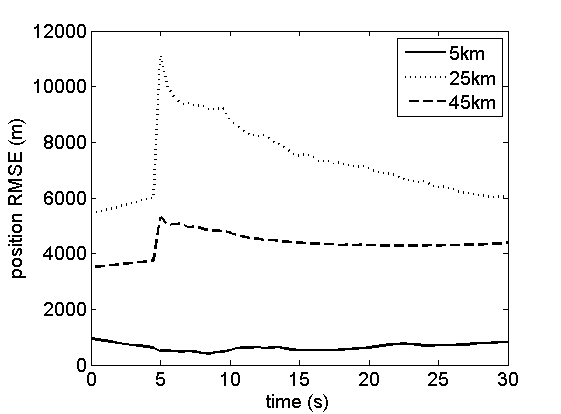}
	\end{center}
	\caption{Position RMSE for the case (b).}
	\label{Fig11}
\end{figure}

The position root mean square error (RMSE) versus time obtained from 100 Monte Carlo runs for cases (a) and (b) is shown in Figs.~\ref{Fig10}--\ref{Fig11}, respectively.
In case (a), the trajectory starting at $45\,\mathrm{km}$ achieves the best estimation accuracy, while the $25\,\mathrm{km}$ trajectory exhibits the worst performance, and the $5\,\mathrm{km}$ trajectory lies in between.
In case (b), the performances of the $5\,\mathrm{km}$ and $45\,\mathrm{km}$ trajectories are reversed, whereas the $25\,\mathrm{km}$ trajectory remains the worst.
These observations are consistent with the observability analysis presented in Section~\ref{s4}, where better observability leads to higher accuracy of the estimation.

An interesting observation is that the batch initial estimate yields better accuracy than the subsequent dynamic estimation when observability is poor, for example in the 25 km scenarios, 5 km in case~(a) and 45 km in case~(b). In the initialization process the initial state $\bd x_0$ is obtained purely from grid searching without invoking the ILS procedure. The measurement residual $\bd Z - \bd g(\bd x_0^0)$ already satisfies the ILS stopping criterion in~(\ref{eqStop}). We then bypassed this stopping criterion and forced the solution to proceed through the ILS iterations. However, the resulting estimate became significantly worse than that from grid searching. This is because, under poor observability conditions, the Jacobian $\bd J$ in~(\ref{eqJac}) becomes ill-conditioned, causing severe numerical issues in the matrix inversion required in~(\ref{eqILS}). Consequently, the ILS fails to produce a reliable estimate based on $\bd J$, whereas the grid searching approach, which directly evaluates the measurement residual without relying on $\bd J$, yields better accuracy. In principle, the ILS formulation is theoretically sound, but in these cases its performance is limited by numerical ill-conditioning. The ILS results are not shown; however, their degradation is consistent with the abrupt increase in the dotted curves around 5 s in Figs.~\ref{Fig10} and~\ref{Fig11}, caused by ill-conditioned dynamic estimation using the same underlying model. The subsequent dynamic estimation using the EKF is affected by the numerical ill-conditioning of the Jacobian, and therefore cannot surpass the accuracy of the initial batch estimate. In any case, the estimation in these scenarios operates under marginal observability; nevertheless, the RMSE study indicates that useful trajectory information can still be recovered.

\section{Conclusions}\label{s6}
This paper studies Doppler-only measurement fusion for two closely spaced stationary Tx--Rx pairs when the target lies within the Tx--Rx baseline ambiguity zone. The observability analysis shows that a target trajectory is observable provided it does not lie along a bistatic ellipse associated with a Tx--Rx sensor pair. Observability is marginal near the middle of the baseline and improves toward both ends, particularly when either the two Rxs or the two Txs at the baseline ends are spatially separated. In addition, higher target speed leads to improved observability.

A Doppler-only measurement fusion approach was developed. A multiple-initial-point ML nonlinear estimator is first employed for initial state estimation, followed by dynamic state updates using an EKF. Simulation results are presented and are consistent with the observability analysis. The position RMSE study indicates that meaningful trajectory information can still be recovered under marginal observability conditions.

The proposed approach applies to both first-time-seen targets within the ambiguity region, using tracking initiation followed by EKF, and targets passing through the region, using EKF only.

Future work will extend the proposed approach to three-dimensional trajectories and investigate alternative dynamic estimation methods that are less sensitive to Jacobian ill-conditioning than the EKF.

%\section*{Acknowledgments}

%The authors would like to thank Mr. Leung Hong Huat Michael for initiating this problem and for providing sufficient materials and support throughout the entire process of this work. His insights and contributions were valuable in shaping our understanding and the direction of this study.

\appendices
\section{Jacobian of $\bd g(\bd x_0)$}\label{a1}
The Jacobian of $\bd g(\bd x_0)$ in~(\ref{eqJac}) is derived here.
For simplicity, each equation in $\bd g(\bd x_0)$ is given below,
omitting the Tx--Rx pair index and iteration index.
\begin{equation}
    \dot r = \dot r\uT + \dot r\uR
\end{equation}
where
\begin{eqnarray}
\dot r\uT &=&
\frac{\Delta x\uT \dot x + \Delta y\uT \dot y}
{r\uT} \\
\dot r\uR &=&
\frac{\Delta x\uR \dot x + \Delta y\uR \dot y}
{r\uR} \\
\Delta x\uT &=& x_0 + t \dot x - x\uT \\
\Delta y\uT &=& y_0 + t \dot y - y\uT \\
\Delta x\uR &=& x_0 + t \dot x - x\uR \\
\Delta y\uR &=& y_0 + t \dot y - y\uR\\
r\uT &=& \sqrt{(\Delta x\uT)^2 + (\Delta y\uT)^2}\\
r\uR &=& \sqrt{(\Delta x\uR)^2 + (\Delta y\uR)^2}
\end{eqnarray}

Its derivatives with respect to $\bd x_0 = [x_0\ y_0\ \dot x\ \dot y]'$ are

\begin{eqnarray}
\frac{\partial \dot r}{\partial x_0}
&=&
\frac{r\uT\,\dot x-\dot r\uT\,\Delta x\uT}{(r\uT)^2}
+
\frac{r\uR\,\dot x-\dot r\uR\,\Delta x\uR}{(r\uR)^2}\\
\frac{\partial \dot r}{\partial y_0}
&=&
\frac{r\uT\,\dot y-\dot r\uT\,\Delta y\uT}{(r\uT)^2}
+
\frac{r\uR\,\dot y-\dot r\uR\,\Delta y\uR}{(r\uR)^2}\\
\frac{\partial \dot r}{\partial \dot x} 
&=&
\frac{r\uT (\Delta x\uT + t \dot x) - t \Delta x\uT \dot r\uT}{(r\uT)^2} \nonumber \\
&&+\frac{r\uR (\Delta x\uR + t \dot x) - t \Delta x\uR \dot r\uR}{(r\uR)^2}\\
\frac{\partial \dot r}{\partial \dot y} 
&=&
\frac{r\uT (\Delta y\uT + t \dot y) - t \Delta y\uT \dot r\uT}{(r\uT)^2} \nonumber \\
&&+
\frac{r\uR (\Delta y\uR + t \dot y) - t \Delta y\uR \dot r\uR}{(r\uR)^2}
\end{eqnarray}
where $t$ is the measurement time interval relative to the time of $\bd x_0$.

The Jacobian $\bd J(\bd x_0)$ is
\begin{equation}
\bd J(\bd x_0) =
\left[
\begin{array}{cccc}
\vdots & \vdots & \vdots & \vdots \\

\frac{\partial \dot r}{\partial x_0} &
\frac{\partial \dot r}{\partial y_0} &
\frac{\partial \dot r}{\partial \dot x} &
\frac{\partial \dot r}{\partial \dot y} \\

\vdots & \vdots & \vdots & \vdots
\end{array}
\right]
\end{equation}

\end{document}